\documentclass[preprint]{aastex}
\usepackage{graphicx}
\usepackage{epsfig}
\usepackage{amsmath}
\usepackage{lineno}
\usepackage{comment}
\usepackage{subfig}
\begin{document}
\shorttitle{}
\shortauthors{Fang et al.}
\title{Buildup of Magnetic Shear and Free Energy 
During Flux Emergence and Cancellation}

\author{Fang Fang\altaffilmark{1}, Ward Manchester IV\altaffilmark{1}, 
  William P. Abbett\altaffilmark{2}, 
  Bart van der Holst\altaffilmark{1} }
\affil{\altaffilmark{1} Department of Atmospheric, Oceanic and Space Sciences, 
  University of Michigan, Ann Arbor, MI 48109, USA}
\affil{\altaffilmark{2} Space Sciences Laboratory, University of California, 
  Berkeley, CA 94720, USA}


\begin{abstract}
We examine a simulation of flux emergence and cancellation, which shows a complex 
sequence of processes that accumulate free magnetic energy in the solar corona 
essential for the eruptive events such as coronal mass ejections (CMEs), filament 
eruptions and flares. The flow velocity at the surface and in the corona shows a 
consistent shearing pattern along the polarity inversion line (PIL), which together 
with the rotation of the magnetic polarities, builds up the magnetic shear. 
Tether-cutting reconnection above the PIL then produces longer sheared magnetic 
field lines that extend higher into the corona, where a sigmoidal structure forms. 
Most significantly, reconnection and upward energy-flux transfer are found to occur 
even as magnetic flux is submerging and appears to cancel at the photosphere. A 
comparison of the simulated coronal field with the corresponding coronal potential 
field graphically shows the development of nonpotential fields during the emergence 
of the magnetic flux and formation of sunspots.

\end{abstract}




\section{Introduction} \label{intro}

Regions of intense solar magnetic fields such as sunspots and active regions
are known to exhibit energetic outbursts that are manifest in many forms, such as 
CMEs, filament eruptions, and flares 
\citep[e.g.][]{forbes2000, low2001, forbes2006, gopalswamy2006, schrijver2009}. 
In the most dramatic form, magnetic fields release energy up to $10^{32}$ ergs, 
and in the process expels $10^{15}$ gm of plasma and $10^{20}$ Mx of magnetic flux 
into interplanetary space \citep{bieber1995} at speeds that can reach 3000 km/s 
\citep[e.g.][]{gosling1990, stcyr2000, zhang2004}. Although less energetic, 
large-scale regions of relatively weak magnetic fields can also erupt giving rise 
to streamer blowouts and the expulsion of quiescent filaments into the solar wind 
\citep[e.g.][]{hundhausen1993, lynch2010}. 

In order to produce such eruptions, the magnetic field must possess free magnetic 
energy, which requires that the field is in a nonpotential configuration with 
electric currents passing through the photosphere and into the corona. In general, 
the field lines of the non-potential force-free fields are stretched, expanded and 
intertwined compared to the equivalent potential field mapping to the photosphere. 
Such force-free fields in the solar corona have historically been taken to be of 
two forms: arcades, in which the foot points have been sheared to produce a strong 
field component parallel to the polarity inversion line (PIL), or flux ropes, in 
which the magnetic field is twisted about a central axis. This dichotomy of field 
configurations naturally leads to two broad groups of theoretical and numerical 
models of CMEs. 

In the case of magnetic arcades, initial configurations are typically taken to be 
line-tied to the lower boundary and subsequently energized with prescribed shear 
flows, which cause the arcade to expand and then erupt 
\citep[e.g.][]{steinolfson1991, mikic1994, amari1996, guo1998}. Another variant, 
the so-called breakout model, involves a system of quadrupolar fields with a system 
of three arcades. The field reverses direction over the central arcade allowing 
magnetic reconnection to release the central arcade 
\citep[e.g.][]{antiochos1999, lynch2004, lynch2008, vanderholst2009}. In all of 
these cases, magnetic flux ropes form during the eruption process and are expelled 
from the corona. For those models, which assume that flux ropes exist prior to the 
eruption, the flux ropes are assumed to reside within an arcade where the 
outward-directed Lorentz force of the rope (hoop force) is balanced by magnetic 
tension of the surrounding arcade \citep[e.g.][]{forbes1991, low1994, titov1999}. 
The central question with flux rope models is how the force imbalance occurs and 
results in eruption. Flux ropes have been energized by the application of rotational 
motions at the rope footpoints \citep[e.g.][]{tokman2002, rachmeler2009} that twist 
up the field. Flux ropes have also been forced to emerge into the corona by bodily 
advecting them through the lower boundary of the computational domain 
\citep[e.g.][]{torok2005, fan2005}.  In the case of \citet{amari2003}, flux 
cancellation at the PIL forms a flux rope within a sheared magnetic arcade. 
Following these formation processes, the flux rope may lose equilibrium and erupt 
from several instabilities including: unbalance of magnetic forces, which is 
followed by magnetic reconnection \citep{forbes1991, amari2003, roussev2004}, 
torus instability \citep{torok2005}, or kink instability \citep{sturrock2001, fan2005}.

Plentiful observations provide evidence for both arcades and flux ropes as the 
coronal field structures for CMEs.  Sheared magnetic fields are measured directly 
at the photosphere with vector magnetographs 
\citep[e.g.][]{hagyard1984, zirin1993, falconer2001, yang2004, liu2005}, where the 
field is found to run nearly parallel to the PIL. More recent analysis by 
\citet{schrijver2005} found that shear flows associated with flux emergence drove 
enhanced flaring. Similarly, active region CME productivity is also strongly 
correlated with magnetic shear, as shown by \citet{falconer2002, falconer2006}.
There is also evidence suggesting that flux ropes may bodily emerge through the 
photosphere to reside in the corona. The twist of the emerging flux tube is 
characterized by the presence of elongated magnetic polarities, i.e., magnetic 
tongues in longitudinal magnetograms \citep{luoni2011}. \citet{leka2001} examined 
a time series of vector magnetograms of emerging magnetic flux and found that 
magnetic fields passing through the photosphere are already twisted. Similarly, 
\citet{lites1995} examined a time series of photospheric vector magnetograms, which 
they found could be fitted by the emergence of a spherically shaped magnetic flux 
rope. \citet{lites2005} shows an example of the photospheric vector magnetic
field and chromospheric structure seen in the H$\alpha$ line, which suggest the
presence of a coronal flux rope associated with a prominence. 

Magnetohydrodynamic (MHD) numerical simulations may provide new insights to 
our understanding of the physical processes that build up the free
magnetic energy necessary for solar eruptions. Global scale simulations 
address how magnetic fields generated at the tachocline pass through the
convection zone and have explained and reproduced many large-scale aspects of 
sunspots \citep[e.g.][]{spruit1982, abbett2001, fan2008}.  However because of 
the anelastic approximation, these models can only extend up to a height of 
20 Mm below the photosphere.  Above this depth, fully compressible MHD models 
have been employed to simulate the emergence of magnetic fields from the 
near surface convection zone into the lower corona.  Early work 
demonstrated the basic processes by which fields emerge to form bipolar
active regions \citep[e.g.][]{shibata1989, matsumoto1993}. These simulations
were followed by cases \citep[e.g.][]{manchester2001, fan2001, magara2003}, 
which showed that the Lorentz force, arising when fields expand into the 
highly stratified solar atmosphere, drives shear flows along the PIL. The shear 
flows then transport magnetic energy into the coronal portion of the flux rope 
which then expands, and drive its eruption \citep{manchester2004}. This 
shearing process is suggested as the energy source for CMEs and flares 
\citep[]{manchester2007, manchester2008}. \citet{galsgaard2007}, 
\citet{archontis2008}, and \citet{mactaggart2009} produce fast CME-like eruptions 
by also invoking magnetic reconnection with emerging flux ropes. \citet{fan2008} 
simulated flux rope emergence that exhibited sunspot rotation driven by a torsional 
form of the Lorentz force, which twisted up the coronal portion of the field as 
predicted by \citet{parker1977}. However, in this case, no eruptive behavior was found.

Recently, computing power has reached a level that has allowed the development 
of realistic solar models, including radiative and thermodynamic processes necessary
to simulate convection in conjunction with the upper atmosphere. 
\citet{stein2006}, \citet{abbett2007}, \citet{rempel2009}, \citet{cheung2010}, 
\citet{kitiashvili2010}, and \citet{rempel2011} 
emphasize the importance of the interaction between magnetic fields and
convective motions.  \citet{stein2006} and \citet{abbett2007}, have 
each addressed quiet-Sun magneto-convection, while \citet{kitiashvili2010} report 
the critical role of strong vortical downdrafts around small magnetic structures 
in the formation of large-scale structures.  The work of \citet{rempel2009} and 
\citet{cheung2010} treats the formation and evolution of sunspots, including 
the Evershed flow. \citet{fang2010, fang2012} address the emergence of 
magnetic flux ropes from a turbulent convection zone into the corona and 
found both shearing and rotational flows driven by the Lorentz force.
These horizontal flows were found to dominate the energy transport from the 
convection zone into the corona.  Of particular significance, the simulation 
of \cite{fang2012} exhibited a case of large-scale flux cancellation, a 
phenomena strongly associated with CME initiation \citep[]{subramanian2001}.  
Here we examine the flux cancellation in conjunction with energy transport from 
the convection zone to the corona.

The paper is organized in the following way:
Section \ref{method} describes the numerical simulation; 
Section \ref{cancel} studies the build up of magnetic shear in
a flux cancellation event;
followed by analysis on the free energy in the corona in Section \ref{corona}.
The energy transfer at the photosphere is discussed in Section \ref{flux}.
Section \ref{conclusion} summarizes our conclusions and discussion.


\section{Numerical Simulation} \label{method}


To simulate the flux emergence in the convection zone, we first 
generate a relaxed solar atmosphere of dimension $30\times30\times42$ Mm$^{3}$,
with the photosphere located at $Z =$ 0 Mm. Our model solves the MHD equations
in Block-Adaptive Tree Solar-wind Roe Upwind Scheme (BAT-S-RUS)
with additional energy source terms to describe thermodynamic processes in 
the solar atmosphere, i.e., radiative cooling at the photosphere and in the 
corona, and coronal heating \citep[]{abbett2007,fang2010}. Meanwhile,
implementation of the non-ideal tabular equation-of-state \citep[]{rogers2000}
provides a more accurate description of the partially ionized plasma 
in the convection zone.  We use periodic horizontal,  closed upper boundary 
conditions, and fix the density and temperature values at the lower boundary 
while setting the vertical momentum to be reflective. 

Taking advantage of the adaptive grid in BAT-S-RUS, our model produces a relaxed 
convection zone with a depth of 20 Mm, with a overlying photosphere of 0.45 Mm 
thickness and a corona of 20 Mm, in a Cartesian domain. The vertical statification 
of the atmosphere in our simulation domain is shown in Figure 1 in \citet{fang2012}. 
We then carry out a simulation of the emergence of a buoyant, initially stationary, 
horizontal flux rope inserted at $Z = -10$ Mm, shown in Figure 2 in \citet{fang2012}. 
Interaction of the rising flux rope with large-scale convective motion produces the 
bipolar structure of the flux rope, with the convective downflows fixing the two 
ends of the emerged section of the flux rope in the convection zone, illustrated in 
Figure 5 in \citet{fang2012}. Near-surface small-scale convection produces the 
magnetic polarities by coalescence and intensifies the strength of the magnetic flux. 
A small active region forms on the photosphere, seen in Figure 6 of 
\citet{fang2012}, exhibiting strong interaction of the magnetic flux and the surface
flows, i.e., the shearing and converging motions, separation and rotation of the 
magnetic polarities. In the presence of these flows, a flux cancellation event
takes place at time t = 05:00:00, shown by Figure 6 of \citet{fang2012}, during 
which 10\% of the total photospheric unsigned flux is cancelled. 
During the flux emergence, horizontal motions dominate the energy transfer 
to the corona, while vertical flows transport energy back into the convection 
zone. Details on the dynamics of the emerging flux rope are discussed 
in \citet{fang2012}.


\section{A case of Flux Cancellation} \label{cancel}


\begin{figure*}[ht!]
  \begin{minipage}[t] {1.0\linewidth}
    \begin{center}
      \subfloat{\label{bzuxy1}\includegraphics[width=160mm]{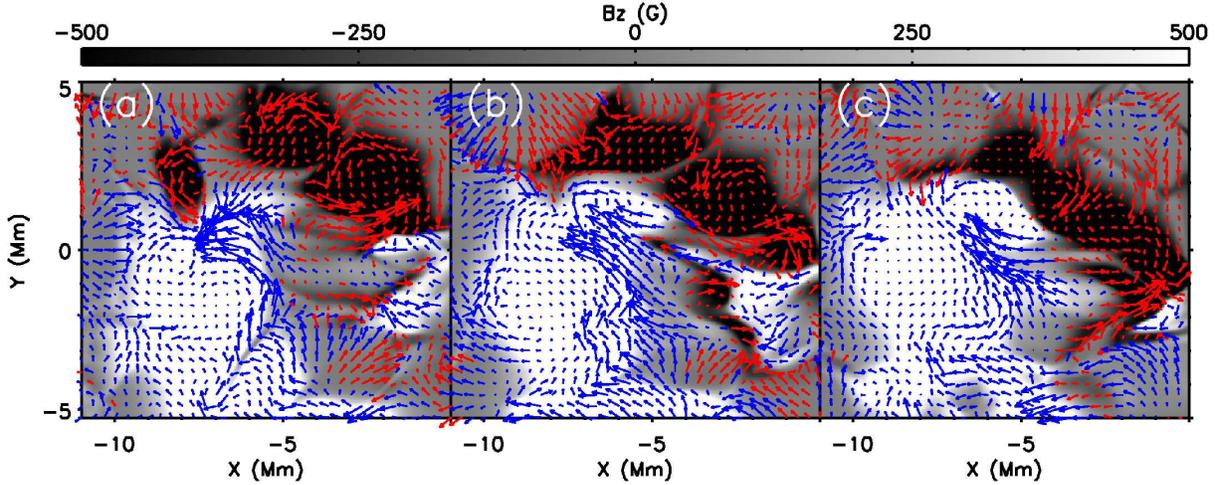}}
      \subfloat{\label{bzuxy2}}
      \subfloat{\label{bzuxy3}}
    \end{center}
  \end{minipage}\hfill
  \caption{The photospheric structure of B$_{z}$ in the area of flux cancellation 
    at time t = 5:00:00 (a), 5:10:00 (b), 5:22:00 (c). Blue and red arrows show 
    the horizontal velocity in positive and negative magnetic polarities, respectively. 
    Shear flows are most apparent at time t = 5:10:00.}
\end{figure*}

\begin{figure*}[ht!]
  \begin{minipage}[t] {1.0\linewidth}
    \begin{center}
      \includegraphics[width=160mm]{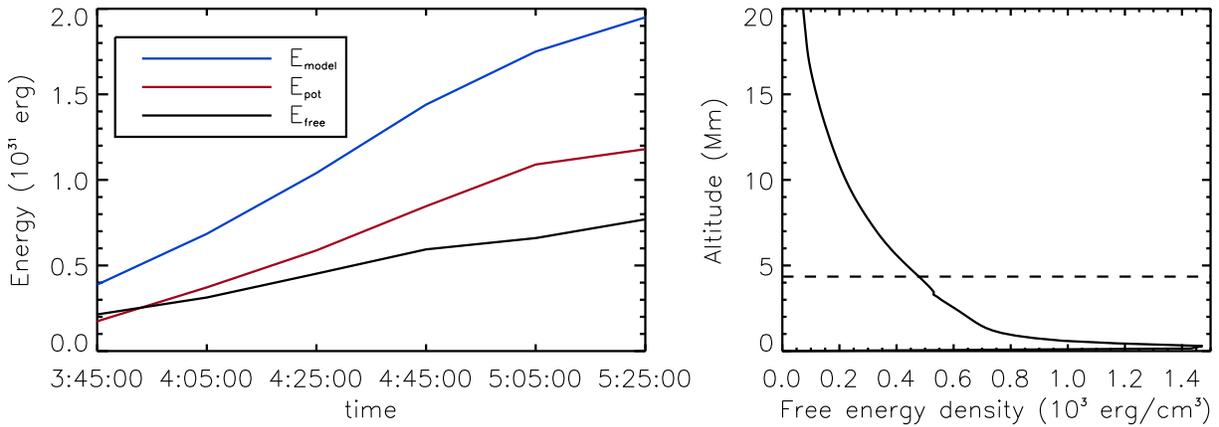}
    \end{center}
  \end{minipage}\hfill
  \caption{Left: temporal evolution of the total (blue), potential (red) 
    and free (black) magnetic energy. Right: vertical distribution 
    of free energy density at time t = 05:05:00. The dashed line indicates
    the altitude up to where 50\% of the coronal free energy is stored.}
  \label{component}
\end{figure*}

Here, we examine a region of magnetic flux cancellation at the photosphere in our 
simulation, which is also associated by high magnetic shear. Flux cancellation of 
this form is particularly significant as it has been observed to be associated 
with flares \citep[]{martin1985} and CMEs \citep[]{subramanian2001} in active 
regions.  A clear example is observed in AR 10977, where reconnection at the PIL 
during flux cancellation forms a highly sheared arcade and sigmoidal structure 
that subsequently reforms after eruption \citep[]{green2011}.  \cite{li2004} 
found converging flows consistent with flux cancellation in decaying and young 
active regions producing CMEs. Inspired by similar observations, converging motion 
and flux cancellation have been imposed as boundary conditions in MHD simulations 
to produce eruptions \citep[]{linker2003,amari2003}.

As it is observed on the Sun, flux cancellation is very ambiguous as it may be 
caused by submergence of $\Omega$-loops, emergence of U-loops, or magnetic 
cancellation by some dissipative process.  Our simulation provides a unique 
opportunity to fully understand an example of flux cancellation in a way that is not 
possible through observations alone. We find flux cancellation occurs spontaneously 
between opposite polarities that are driven together by the convective flows near 
the photosphere. The temporal evolution of cancellation is shown in Figure 
\ref{bzuxy1}-\ref{bzuxy3}, which show a zoom-in view of the relevant area, with 
background color showing photospheric B$_{z}$ field. The horizontal motions in positive 
and negative polarities are shown by  blue and red arrows, respectively. The flux 
cancellation event starts at time t = 05:00:00, lasts for 0.5 hr, and the total amount 
of cancelled flux approaches $1\times10^{20}$ Mx, which is 10\% of the total unsigned 
flux on the photosphere.  It is remarkable that during the process of flux 
cancellation, the coronal free energy (shown in the left panel of Figure 
\ref{component}) is still increasing even though the photospheric magnetic flux is 
decreasing. 

\begin{figure*}[ht!]
  \begin{minipage}[t] {1.0\linewidth}
    \begin{center}
      \subfloat{\label{z=01}\includegraphics[width=45mm,angle=-90]{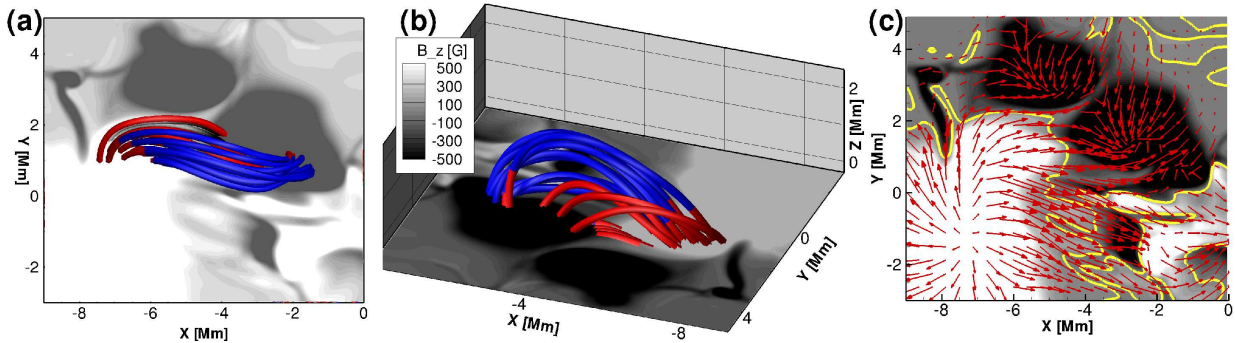}}
      \subfloat{\label{z=02}}
      \subfloat{\label{z=03}}
    \end{center}
  \end{minipage}\hfill
  \caption{The structure of the magnetic fields in the area of flux cancellation
  at time t = 5:05:00 from top (a and c) and side (b) view. The background planes 
  show the photospheric B$_{z}$ field. Color on the rods indicates the vertical 
  velocity at which the field lines are moving, with blue showing rising lines and 
  red submerging. The red arrows in Panel (c) represent the horizontal magnetic 
  field, with yellow line showing PIL.}
\end{figure*}

\begin{figure*}[ht!]
  \begin{minipage}[t] {1.0\linewidth}
    \begin{center}
      \subfloat{\label{lorentz1} \includegraphics[width=160mm]{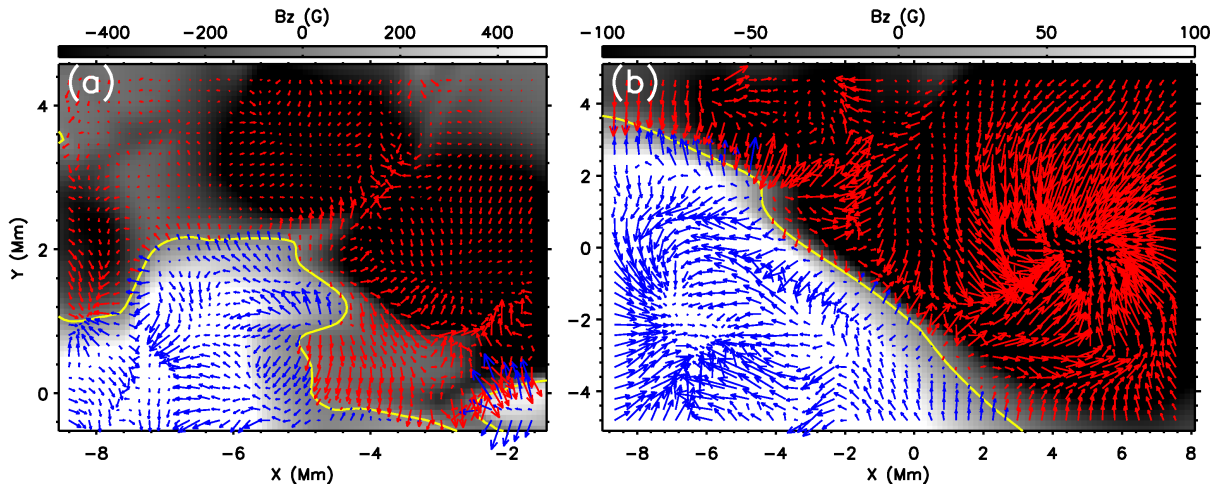}}
      \subfloat{\label{lorentz2}}
    \end{center}
  \end{minipage}\hfill
  \caption{The structure of horizontal Lorentz force on Z = 0 (a) and Z = 3 Mm (b) 
    planes at times t = 05:01:00 (a) and 05:05:00 (b), respectively.
    The blue and red arrows represent Lorentz forces in positive and negative 
    polarities, respectively. The magnitude of $B_{z}$ field component is shown  
    in gray scale.  Note that the Lorentz force reverses direction across the
    polarity inversion line driving the shear flow, and forms a spiral 
    pattern in the negative polarity driving a rotating flow.}
\end{figure*}

To understand this process, we take a close examination of the velocity fields
in the area of flux cancellation prior to, during and in the later phase of 
the event, as shown by Figure \ref{bzuxy1}, \ref{bzuxy2} and \ref{bzuxy3},
respectively. In Figure \ref{bzuxy1}, the two polarities are pushed together by the 
horizontal converging motion at the two sides of the PIL.  The flow pattern 
at the PIL is mostly converging and forms a very narrow PIL at a later time 
illustrated in Figure \ref{bzuxy2}. Figure \ref{bzuxy2} shows a clear shearing motion at the PIL, 
with two polarities running in opposite directions. 
The strong shear flow at the PIL is driven by the Lorentz force, and 
takes place during the flux cancellation as the magnetic field expands
in the corona as shown by the blue colored field lines in Figure \ref{z=01} and \ref{z=02}.
Figure \ref{lorentz1} shows the horizontal Lorentz force at time 
t = 05:01:00 with arrows, from which we find a clear pattern of the force 
running in the opposite directions across the PIL in the area of flux cancellation.
At time t = 5:22:00, toward the end of the flux cancellation, the shearing motion 
is weakened and the flow pattern becomes dominated by the converging motion again, as 
shown in Figure \ref{bzuxy3}. 
The weak-strong-weak shearing motion in the flux 
cancellation event is also shown by observations \citep[]{su2007} during 
flares where footprints of the magnetic fields present the same pattern of motion. 

Figure \ref{z=01} - \ref{z=03} show the configuration of the magnetic fields in 
the area of flux cancellation at time t = 05:05:00. Figure \ref{z=01} provides 
the top view of the fields, and we find that instead of short loops perpendicular 
to the PIL, as is the case for potential fields, the field lines are instead 
elongated parallel to the PIL. Furthermore, the fields are more parallel to the 
PIL as they become closer to it, as shown by the red arrows in Figure \ref{z=03}. 
At the photosphere, the magnetic fields are submerging to the convection zone at 
the PIL, as shown by the red rods in Figure \ref{z=02}. The concentration of both 
the magnetic and velocity shear along the PIL demonstrates that the elongation of 
the field lines is the result of shearing motion, which is further enhanced by the 
converging motion at the PIL \citep[]{martens2001}. These converging motions also 
gives rise to occurrence of tether-cutting reconnection close to the photosphere 
in the highly sheared fields as proposed by \citet{moore2001} as an explanation 
for solar eruptive events. Figure 1 of \citet{moore2001} envisioned reconnection 
inside a highly sheared core field, which would produce longer, nearly horizontal 
field lines higher in the corona, while at the base of the arcade, a system of 
short unsheared loops would form. This flux transfer process accumulates the free 
energy of the sheared field higher in the corona while at the same time reducing 
the magnetic tension by detaching the system from the photosphere. The rise of the 
arcade naturally leads to necking off of the fields, which in turn, promotes more 
reconnection. Unabated, the runaway reconnection provides a mechanism for the 
initialization and growth of explosive events. With our simulation, we can fully 
examine the way in which submergence and reconnection alter the field line geometry 
in the tether-cutting process. A three-dimensional (3-D) view of the magnetic fields 
combined with the flow pattern provides a complete picture of the effects of the 
internal reconnection. Figure \ref{z=02} illustrates the structure of the magnetic 
field lines colored by the vertical component of the velocity orthogonal to the 
field lines. Blue indicates upflows and red downflows. We find two groups of 
magnetic field lines formed during tether-cutting reconnection, with one rising up 
into the corona and the other, shorter loops, submerging into the convection zone. 
At the photosphere, the submergence of the shorter loops, as shown by the red rods 
right above the PIL, results in the decreased unsigned flux observed at the surface. 
Along the long, rising loops, magnetic shear accumulates, forming a highly sheared 
arcade.

\begin{figure*}[ht!]
  \begin{minipage}[t] {1.0\linewidth}
    \begin{center}
      \subfloat{\label{bzflux1}}
      \subfloat{\label{bzflux2}}
      \subfloat{\label{bzflux3}}
      \subfloat{\label{bzflux4}}
      \subfloat{\label{bzflux5}}
      \subfloat{\label{bzflux6} \includegraphics[width=160mm]{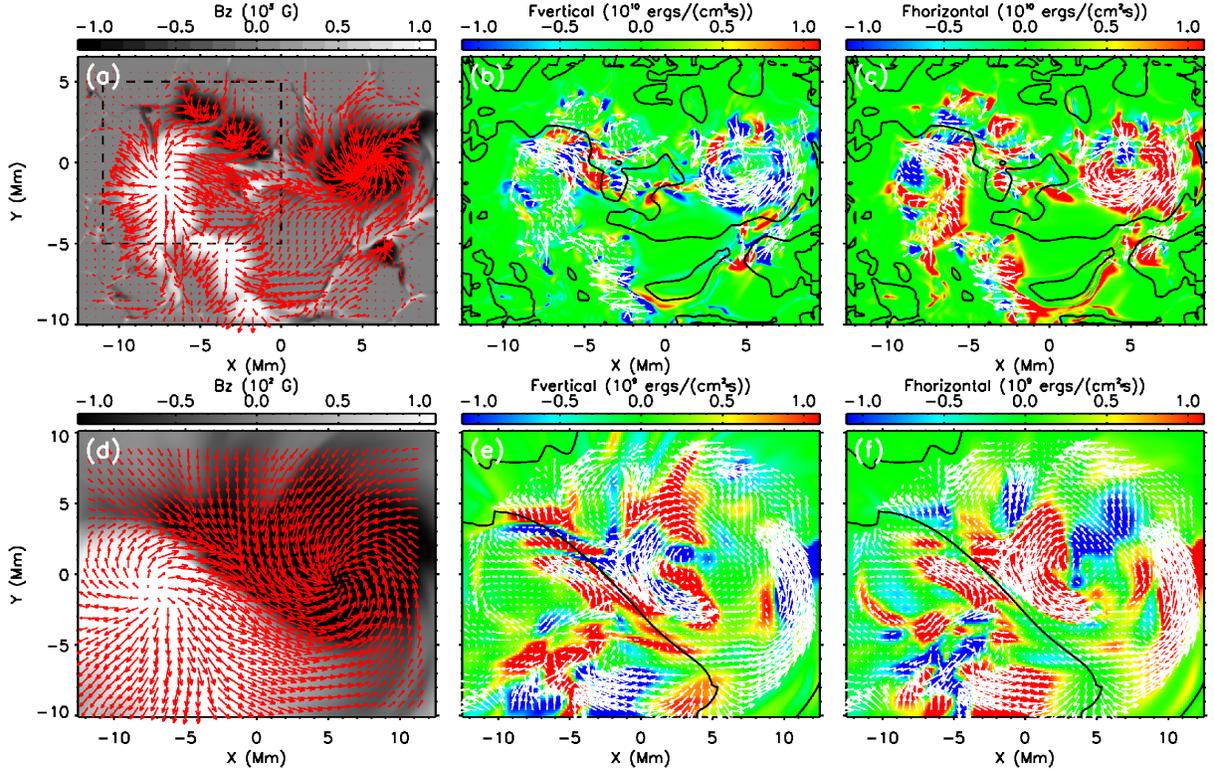}}
    \end{center}
  \end{minipage}\hfill
  \caption{The structure of B$_{z}$ (a and d), F$_{\mathrm{vertical}}$ (b and e) 
    and F$_{\mathrm{emerge}}$ (c and f) fields
    at the photosphere (a, b and c) and at Z = 3 Mm in the corona 
    (d, e, and f) at time t = 05:05:00. PIL is shown by the black line.
    Red arrows in Panel (a) and (d) represent the horizontal magnetic fields,
    and white arrows in (b, c, e and f) show the horizontal velocity fields.}
\end{figure*}

To study the energy transport into the corona during flux cancellation, we examine 
the Poynting flux at $Z= 3$ Mm associated with the vertical motions. At the surface, 
the vertical motion transports the energy back into the convection zone due to the 
concentration of the magnetic flux in downflows. In the corona, the acceleration of 
magnetic fields after reconnection plays a very important role in the energy flux, 
shown by Figure \ref{bzflux5}. At the PIL in the corona, we observe a strong energy 
input associated with the rising motion of the fields.  The rising motion here in 
the corona, however, is different from the emerging motion at the photosphere. At 
the photosphere, the emergence is caused by the expansion and buoyancy of the flux 
rope as well as convective upflows. However, in the corona, the tether-cutting 
reconnection forms longer loops possessing less magnetic tension, resulting in a 
magnetic pressure gradient force accelerating the magnetic fields upward. 

\begin{figure*}[ht!]
  \begin{minipage}[t] {1.0\linewidth}
    \begin{center}
      \includegraphics[width=90mm,angle=-90]{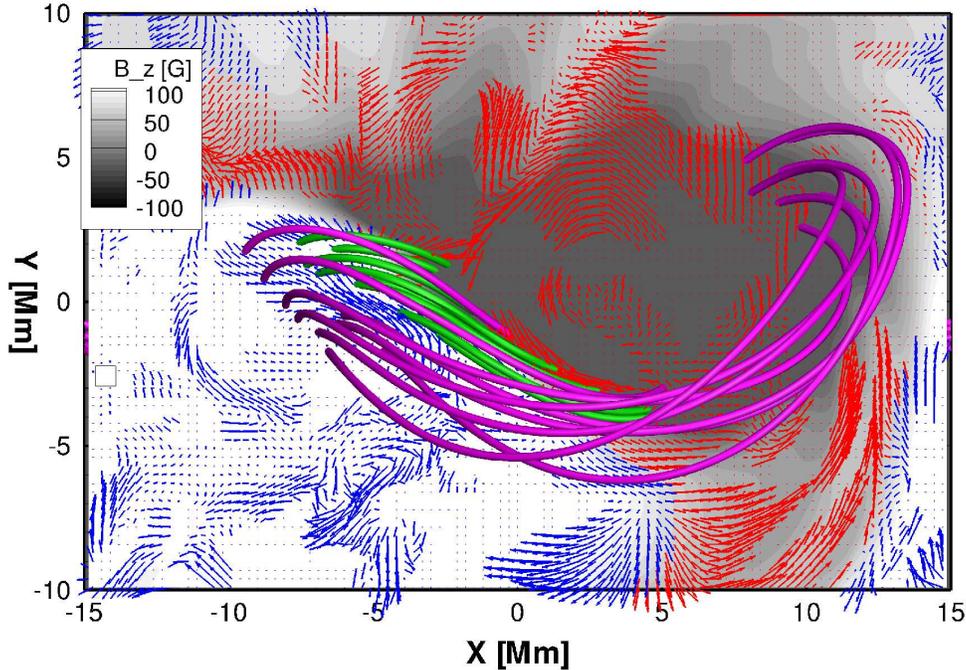}
    \end{center}
  \end{minipage}\hfill
  \caption{3-D structure of the magnetic fields in corona at time t = 5:05:00.
    The background planes show the B$_{z}$ field at Z = 3 Mm.
    Green rods represent low-lying field lines below 5 Mm with purple showing
    high field lines. 
    The blue and red arrows show the horizontal velocity in positive and 
    negative magnetic polarities, respectively. There is a clear shear flow
    along the polarity inversion line.  The negative polarity also shows some signs 
    of rotation, while the positive polarity shows only a coherent shear flow.
    The magnetic field lines clearly reflect the flow pattern.}
  \label{z=3}
\end{figure*}

Besides reconnection and shear flows, rotation of the magnetic polarities also 
contributes to the accumulation of magnetic shear by moving the footpoints in 
opposite directions at the PIL. Figure \ref{z=3} shows the geometry of two groups 
of coronal magnetic field lines: purple overlying lines with apex higher than 5 Mm, 
and green lines lower than 5 Mm. Horizontal velocity fields at $Z = 3$ Mm, 
represented by the arrows, show a rotation pattern on the negative polarity, which 
is also seen in Figure \ref{bzflux5}.  The rotation complements the shearing motion 
at the PIL, in the sense that both shear and rotation contribute to plasma motions 
parallel to the PIL.  However the flow patterns are also quite distinct occurring 
at different scales and at different times. The shear flows are localized to the 
PIL and persist even when there is not a coherent rotation pattern in the magnetic 
polarities. When rotation is present, the shear flow tends to be greater in 
magnitude and will be shown graphically to dominate the energy transport into the 
corona.  

The combination of the two motions creates a highly sheared arcade at the PIL, 
shown by the green rods in Figure \ref{z=3}, and rotation produces the twisted 
field structure at the far ends of the emerged fields, shown by purple rods. 
Both motions are driven by Lorentz force, as shown by the arrows in Figure 
\ref{lorentz2}. At the PIL, the horizontal Lorentz force runs in the opposite 
direction, consistent with the directions of the shearing motion. In the negative 
polarity, the Lorentz force rotates in the same way of the rotation of the polarity. 
The sigmoidal structure at the PIL is maintained by the consistent shearing and 
rotating motions in our simulation. The relative contribution of shear flows 
compared to rotation can be seen in the geometry of the coronal fields at the PIL.  
Here, the field is dominated by an arcade structure instead of a twisted flux rope. 
Most of the twist is located at the outer periphery of the two polarities. Even in 
the most twisted field lines, there is less than a full turn of the field around 
the axis of either polarity. The absence of the twisted flux rope may be explained 
by two facts: the axis of the initial magnetic flux rope remains in the convection 
zone; and there are not enough twisting motions on the surface to reproduce such a 
twisted structure after emergence. The emerged fields are sheared and elongated, 
forming the arcade structure over the PIL.  


\section{Non-potentiality of the Magnetic Field} \label{corona}


\begin{figure*}[ht!]
  \begin{minipage}[t] {1.0\linewidth}
    \begin{center}
      \includegraphics[width=160mm]{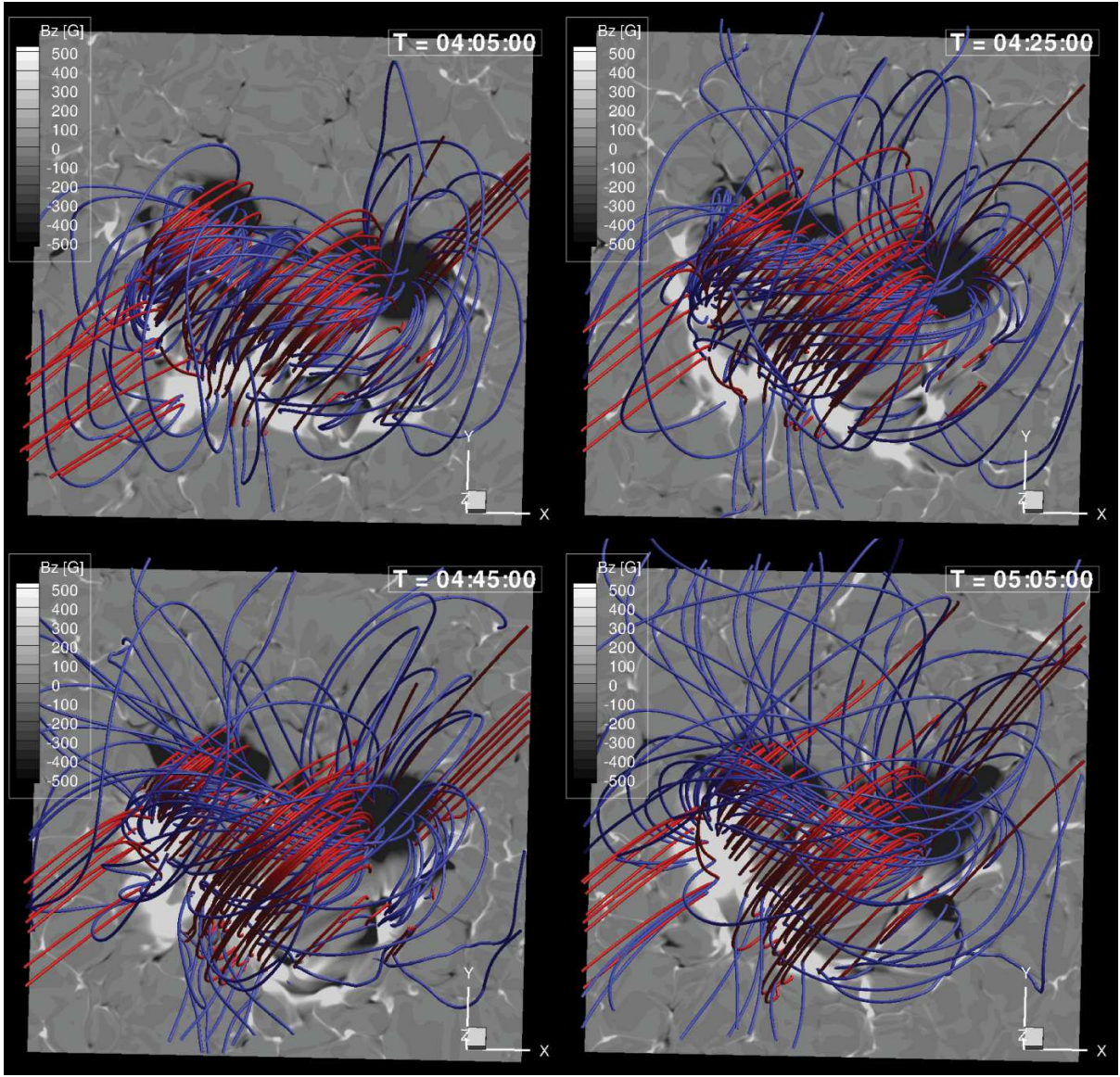}
    \end{center}
  \end{minipage}\hfill
  \caption{Comparison of the model field (blue) with the extrapolated 
    potential field (red) at time t = 04:05:00, 04:25:00, 04:45:00 and 
    05:05:00. The color on the plane shows the B$_{z}$ field 
    at the photosphere.  The clearest departure in field line direction
    occurs at the sheared field lines above the PIL.}
  \label{pfmhd}
\end{figure*}

During our simulation, a significant amount of energy, up to 2$\times10^{31}$ ergs, 
is transferred into the corona, over 5 hours of evolution. The question is then, 
how much of the magnetic energy is in a free form, which is necessary for solar 
eruptive events. To investigate the evolution of magnetic free energy, we compare 
the coronal magnetic fields in our simulation with potential fields extrapolated 
from the photospheric boundary conditions. Figure \ref{pfmhd} shows the temporal 
evolution of the model fields (blue) and the extrapolated potential fields (red). 
Comparison between the two fields clearly shows the buildup of magnetic shear 
along the field lines during the process of flux emergence. At time t = 04:05:00, 
shown by the upper-left panel in Figure \ref{pfmhd}, the magnetic flux concentrates 
in two polarities, with most of the field lines running perpendicular to the PIL. 
The energy transfer into the corona occurs as the magnetic field lines become 
elongated and sheared along the PIL.  The lower-right panel in Figure \ref{pfmhd} 
shows the field structure at time t =05:05:00, in which the field lines are 
sheared along the PIL and  compressed into the lower atmosphere while the 
overlying fields remain almost potential. The simulated configuration of the 
magnetic fields in our domain is consistent with observations and simulations 
of the magnetic fields before solar eruptions \citep[]{schrijver2008,leake2010}. 
The overlying fields confine and compress the sheared core in the lower atmosphere, 
which may be destabilized and give rise to sudden release of magnetic free energy. 

\begin{figure*}[ht!]
  \begin{minipage}[t] {1.0\linewidth}
    \begin{center}
      \subfloat{\label{alpha1}}
      \subfloat{\label{alpha2}\includegraphics[width=60mm,angle=-90]{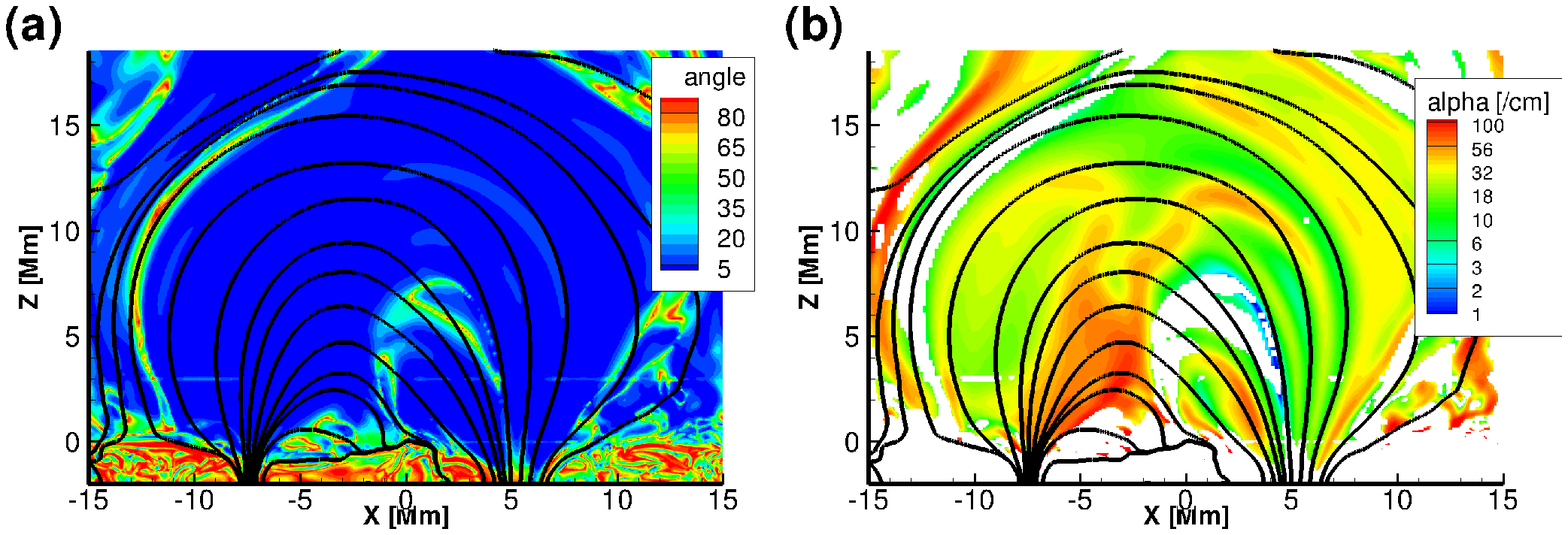}}
    \end{center}
  \end{minipage}\hfill
  \caption{(a) the angle between the current and magnetic fields (b)
    the $\alpha$ structure on the Y = 0 plane. Black lines are the magnetic
  field lines and white regions in Panel (b) are areas where the angle 
  between the current and magnetic fields are greater than $10^{\circ}$.}
\end{figure*}

Here, we calculate the free energy for the simulation by integrating the magnetic 
energy for the model and subtracting the energy of the potential field. The results 
are shown in the left panel of Figure \ref{component} where the temporal evolution 
of model, potential and free energy is plotted with blue, red and black lines, 
respectively. Beginning at time t = 04:05:00, the amount of free energy doubles 
within one hour (the period shown by Figure \ref{pfmhd}), such that the coronal 
free energy approaches $7\times10^{30}$ ergs while the total energy is 
$1.7\times10^{31}$ ergs. At this time, the free energy contributes to 40\% of the 
total energy. Meanwhile, the vertical stratification of the free magnetic energy, 
is shown in the right panel of Figure \ref{component}.  The line plot shows a strong 
tendency of concentrating free energy in the lower atmosphere, with 50\% located 
below $Z = 4$ Mm, shown by the dashed line.  Figure \ref{alpha1} and \ref{alpha2} 
shows the angle between the current and the magnetic field and the structure of 
$\alpha$ (the ratio of $|4\pi\mathbf{j}|$ and $|\mathbf{B}|$ when the angle between 
them is less than 10$^{\circ}$) in the Y = 0 plane, respectively. The non-force-free 
magnetic fields, as shown in Figure \ref{alpha1} mainly reside in the convection zone, 
the core sheared field close to the surface and the negative polarity where the 
rotation forms a twisted magnetic structure (see Figure \ref{pfmhd}). Note that 
$\alpha$ tends to be maximized low down the center of the arcade structure and 
then falls off with distance from the PIL.  


\section{Poynting Fluxes} \label{flux}


To separate the energy transfer associated with vertical and horizontal flows, we 
decompose the Poynting flux into two components: 
\begin{eqnarray}
  F_{\mathrm{horizontal}}& = &- \frac{1}{4\pi}\left(B_{x}u_{x} +  B_{y}u_{y}\right)B_{z} ,\\
  F_{\mathrm{vertical}}& = & \frac{1}{4\pi}\left(B_{x}^{2} + B_{y}^{2}\right)u_{z} .
\end{eqnarray}
F$_{\mathrm{vertical}}$ indicates the energy transported by the vertical motion, while 
F$_{\mathrm{horizontal}}$ represents the energy flux due to the horizontal motions, 
which include, at the photosphere, the separation of small dipoles, rotation, and 
the shearing flow along the PIL. Figure \ref{bzflux2} and \ref{bzflux3} show the 
structures of F$_{\mathrm{vertical}}$ and F$_{\mathrm{horizontal}}$, respectively, with 
the arrows representing the horizontal velocity fields. It is clear that in regions 
with rotation or shearing there is a strong energy input into the corona. Comparison 
of the structures of F$_{\mathrm{vertical}}$ and F$_{\mathrm{horizontal}}$ suggests that 
horizontal motion dominates in the energy transfer during the current phase of 
emergence, while vertical motion transfers energy back into the convection zone, 
due to the concentration of magnetic flux in downflow regions 
\citep[]{fang2010, fang2012}. Vertical emerging motion dominates at the very 
beginning of the flux emergence, and later gives way to horizontal motions, a pattern 
found earlier by \citet{magara2003} and \citet{manchester2004}. 

At the photosphere, the magnetic field is very highly structured by convection. 
In particular, the flow patterns and Poynting fluxes can be difficult to interpret 
with a large component of their structure being caused by random convective motions.  
To get a better understanding of the dynamics, we plot the vertical field strength, 
flow field and Poynting fluxes in the low corona at a height of $Z =3 $ Mm in Figure 
\ref{bzflux4} - \ref{bzflux6}. Here, in the corona, the magnetic field expands and 
smooths out as the plasma beta drops.  This transition allows us to make a much 
clearer picture of the relevant structures.  The flux expands to form two nearby 
polarities and an elongated PIL between them, shown by Figure \ref{bzflux4}. Here, 
the presence of a highly sheared field is obvious in the red arrows showing the 
horizontal field direction. The positive polarity shows little discernible twist 
while the negative polarity is clearly twisted. Flows are completely dominated by 
shearing motion at the PIL and rotation in the polarities. Both flows make a 
significant contribution to the energy transfer, but the energy transported by 
shear flow tends to be more consistent in time than that of rotational flows. 


\section{Summary and Conclusions} \label{conclusion}


Our simulation provides us with a unique opportunity to study an example of 
magnetic flux emergence in a realistic simulation of the convection zone, 
coronal system. The system shows a wealth of complexity and interaction of
mutiple physical processes, which conspire to transport magnetic flux
and free energy from the convection zone into the corona. 
The emerging magnetic flux interacts with convective cells of varying scales
as it approaches the photosphere, and then accumulates in convective downdrafts 
to form a bipolar magnetic structure resembling a solar active region.
The magnetic field continually expands into the upper atmosphere, which results 
in shear flows and rotating motions that combine to draw the field nearly 
parallel to the PIL, and transport energy into the corona.
At a later time, converging motions at the PIL cause flux cancellation at the 
photosphere along with tether-cutting reconnection, which produces 
highly sheared, sigmoidal shaped field lines high in the corona. 

In the process of building up the magnetic free energy, shearing flows play 
a significant role at the photosphere and in the corona. At the near surface 
layers, the angle between the current and magnetic fields ranges from 
$35^{\circ}$ to $90^{\circ}$. The geometry of current and magnetic fields then 
produces the Lorentz force at the surface, which drives the rotating and 
shearing motions and increases the non-potentiality of the magnetic fields 
over a period of hours after the emergence. In particular, the shearing flows 
align the magnetic fields nearly parallel to the PIL and transport significant 
amount of magnetic energy from convection zone into the corona during the flux 
emergence, consistent with observations by \citet{schrijver2005}. The shearing 
flows at the PIL are accompanied by periods of rotating motion of the magnetic 
polarities. 
Both of the two motions contribute to the build up of magnetic shear and free 
energy in the corona. However, the shearing motions distinguish themselves 
from the rotation by their persistent presence and concentration at the PIL where 
the field has an arcade structure, while rotation on the other hand is only present 
in one of the magnetic polarities and lasts only for 1 hour. 
It is the long-lasting, Lorentz-force driven shearing motion that dominates the 
energy transfer.

Convection-driven converging motion drives flux bundles of opposite polarities 
together, producing strong magnetic gradient and the pronounced PIL, preferential 
for large flares \citep{schrijver2007,falconer2008}. The horizontal converging 
motion at the PIL also contributes to the development of the magnetic shear of 
the field lines by increasing the shear angle \citep{martens2001}. The combination 
of converging and shearing flows at the PIL forms a group of highly sheared arcades 
overlying a sharp magnetic PIL. Within this arcade, converging motions lead to the 
occurrence of tether-cutting reconnection, producing two types of field lines: one 
long sheared expanding loops and the other unsheared submerging loops. The short 
loops sink into the convection zone at the speed of convective flows, up to 2 km/s, 
consistent with \citet{harvey1999}, which continuously reduces the photospheric 
unsigned flux. The longer loops rise into the corona with more magnetic shear.
The shearing motion along the PIL plays a very important role in that it both 
produces a sheared arcade structure ready for reconnection and accumulates magnetic 
shear in the field lines formed after the reconnection. The magnetic configuration 
of this area thus yields a high free energy up to 40\% of the total, which is 
comparable with the threshold value for eruptions reported by \citet{amari2003}, 
\citet{aulanier2010} and \citet{moore2012}.

Shear flows, converging motions, and tether-cutting reconnection combine to 
continuously build up the magnetic shear and free energy in the corona necessary 
for eruptive and explosive events. The magnetic reconnection and shearing motion at 
the photosphere produce the sigmoid-shaped field geometry, which is preferential for 
CMEs \citep[]{canfield1999}, and the persistent sheared arcade structure is consistent 
with many CME models 
\citep[e.g.][]{steinolfson1991, mikic1994, amari1996, guo1998, antiochos1999}. Futhermore, 
we find coronal free energy grows at a rate of $3\times10^{30}$ ergs/hr, which builds up 
$10^{31}$ ergs free energy over 5 hours. Such a build up rate is observed by 
\citet{schrijver2007} prior to large flares. Moreover, the majority of the free energy 
resides low in the corona where it can be confined and later released by reconnection. 
Finally, we note that the consistent shearing motion and reconnection at the flux 
cancellation site keep reforming the sigmoid structure, which is essential for homologous 
eruptive events.

In our simulation here, tether-cutting reconnection and flux cancellation take place 
in the highly sheared magnetic fields along the PIL, in the presence of converging and 
shearing flows. In the coupled system of the convection zone, the photosphere and the 
corona, all these mechanisms combine to work simultaneously and build up the free 
magnetic energy in the coronal fields. More importantly, the resulting geometry of 
the magnetic fields in our domain consists of compact, highly sheared core fields 
confined by more relaxed, overlying fields. 50\% of the corona free magnetic energy 
is stored within the compressed, sheared core, while the overlying fields relax very 
quickly with altitude. The compact core, possessing free magnetic energy up to 
10$^{30}$ ergs, provides the preferential magnetic structure as well as free energy 
for solar eruptions. We expect that in future simulations of larger scales, shear 
flows, converging motion and reconnection will produce enough free magnetic energy 
and give rise to eruptions of the magnetic system.

\acknowledgments

The authors thank Spiro K. Antiochos and Tamas I. Gombosi for constructive 
discussions and comments. This work was supported by NASA grant NNG06GD62G, 
NNX07AC16G and NSF grant ATM 0642309 and AGS 1023735. W. M. IV was also funded by 
NASA grant LWS NNX09AJ78G. W. P. A. was supported in part by NASA LWS TR$\&$T award 
NNX08AQ30G, and the Heliophysics Theory Program, under NASA grant NNX08AI56G-04/11. 
The simulations described here were carried out on the Pleiades system at the NASA 
Advanced Supercomputing (NAS) Facility and Bluefire cluster at NCAR. 

\bibliographystyle{apj}
\bibliography{ref}

\end{document}